%% file: clumpy3-arxiv.tex
\documentclass[final,5p,times,twocolumn]{elsarticle}
\usepackage{amsmath}
\usepackage[normalem]{ulem}
\usepackage{color}
\usepackage{colortbl}
\usepackage{units} 
\usepackage{siunitx}
\usepackage[super]{nth} 
\usepackage{bm} 
\usepackage{footmisc}

\usepackage[numbers]{natbib}
\usepackage[pdftex,             
            breaklinks=true,%
            colorlinks=true,%
            citecolor=black,
            pdfauthor={Hütten et al.},%
            pdftitle={CLUMPY: third release}%
           ]{hyperref}

\newcommand{\gr}{$\gamma$-ray}
\newcommand{\grs}{$\gamma$-rays}
\newcommand{\jfactor}{{$J$-factor}}
\newcommand{\jfactors}{{$J$-factors}}

\newcommand{\secref}[1]{\S\ref{#1}}
\newcommand{\eq}[1]{Eq.~(\ref{#1})}
\newcommand{\eqpar}[1]{(Eq.~\ref{#1})}
\newcommand{\eqs}[2]{Eqs.~(\ref{#1}) and (\ref{#2})}
\newcommand{\fig}[1]{Fig.~\ref{#1}}
\newcommand{\figs}[2]{Figs.~\ref{#1} to \ref{#2}}
\newcommand{\tab}[1]{Tab.~\ref{#1}}
\newcommand{\beq}{\begin{equation}}
\newcommand{\eeq}{\end{equation}}
\newcommand{\balign}{\begin{align}}
\newcommand{\ealign}{\end{align}}

\newcommand{\lcdm}{{\ifmmode \Lambda{\rm CDM} \else $\Lambda{\rm CDM}$\fi}}

\newcommand{\nside}{{N_{\rm side}}}

\newcommand{\sigmav}{\ensuremath{\langle \sigma v\rangle}}

\newcommand{\rhosub}{\ensuremath{\overline{\rho}_{\rm subs}}}
\newcommand{\rhotot}{\ensuremath{\overline{\rho}_{\rm tot}}}

\newcommand{\dd}{\ensuremath{\mathrm{d}}}

\newcommand{\planck}{{\it Planck}}
\newcommand{\clumpy}{{\tt CLUMPY}}
\newcommand{\cfitsio}{{\tt cfitsio}}
\newcommand{\healpix}{{\tt HEALPix}}
\newcommand{\great}{{\tt GreAT}}
\newcommand{\rootcern}{{\tt ROOT}}
\newcommand{\doxygen}{{\tt Doxygen}}
\newcommand{\fits}{{\tt fits}}
\newcommand{\python}{{\tt Python}}
\newcommand{\gsl}{{\tt GSL}}
\newcommand{\git}{{\tt git}}

\definecolor{lg}{gray}{0.92}
\definecolor{darkgreen}{rgb}{0,0.4,0}
\definecolor{modelref}{rgb}{0.247,0.3647,0.491}
\definecolor{firebrick}{rgb}{0.698,0.1333,0.1333}
\definecolor{seagreen}{rgb}{0.180,0.545,0.341}

\journal{Computer Physics Communications}

\begin{document}

\begin{frontmatter}



\title{\clumpy{} v3: $\gamma$-ray and $\nu$ signals from dark matter at all scales}

\author[label1,label2,label3,label4]{Moritz H\"utten}
\ead{mhuetten@mpp.mpg.de}
\author[label2]{C\'eline Combet}
\ead{celine.combet@lpsc.in2p3.fr}
\author[label2]{David Maurin}
\ead{dmaurin@lpsc.in2p3.fr}

\address[label1]{Max-Planck-Institut f\"{u}r Physik, F\"{o}hringer Ring 6, D-80805 M\"{u}nchen, Germany}
\address[label2]{Laboratoire de Physique Subatomique et de Cosmologie, Universit\'e Grenoble-Alpes,
CNRS/IN2P3, 53 avenue des Martyrs, 38026 Grenoble, France}
\address[label3]{Deutsches Elektronen-Synchrotron, Platanenallee 6, D-15738 Zeuthen, Germany}
\address[label4]{Humboldt-Universit\"{a}t zu Berlin, Newtonstra{\ss}e 15, D-12489 Berlin, Germany}

\begin{abstract}
We present the third release of the \clumpy{} code for calculating
$\gamma$-ray and $\nu$ signals from annihilations or decays in dark
matter structures. This version includes the mean extragalactic signal
with several pre-defined options and keywords related to cosmological
parameters, mass functions for the dark matter structures, and
$\gamma$-ray absorption up to high redshift. For more flexibility and
consistency, dark matter halo masses and concentrations are now defined with respect to a user-defined overdensity $\Delta$. We have also made changes for the user's benefit: distribution and versioning of the code via \git{}, less dependencies and a simplified installation, better handling of options in run command lines, consistent naming of parameters, and a new Sphinx documentation at \url{http://lpsc.in2p3.fr/clumpy/}.
\end{abstract}

\end{frontmatter}
%
%
{\bf PROGRAM SUMMARY}

\begin{small}
\noindent
{\em Program Title:} \clumpy{}                                   \\
{\em Programming language:} C/C++                           \\
{\em Computer:} PC and Mac                                         \\
{\em Operating system:} UNIX(Linux), MacOS X                    \\
{\em RAM:} between \unit[500]{MB} and \unit[10]{GB} depending on the size or resolution of requested 2D skymaps        
{\em Classification:} 1.1, 1.7, 1.9               \\
{\em External routines/libraries:} 
\gsl{} (\url{http://www.gnu.org/software/gsl}),
\cfitsio{} (\url{http://heasarc.gsfc.nasa.gov/fitsio/fitsio.html}),
CERN \rootcern{} (\url{http://root.cern.ch}; optional, for interactive figures and stochastic simulation of halo substructures),
\great{} (\url{http://lpsc.in2p3.fr/great}; optional, for MCMC Jeans analyses) \\
{\em Nature of problem:} Calculation of the  $\gamma$-ray and $\nu$ signals from dark matter annihilation/decay at any redshift $z$.
\\
{\em Solution method:} New in this release: Numerical  integration of moments (in redshift and mass) of the mass function, absorption, and intensity multiplier (related to the DM density along the line of sight).
   \\
{\em Restrictions:}
Secondary radiation from dark matter leptons, which depends on astrophysical ingredients (radiation fields in the Universe) is the last missing piece to provide a full description of the expected signal.
   \\
{\em Running time:} Depending on user-defined choices, primarily on  substructure modelling and numeric precision $\epsilon$: For instance, it takes \unit[$\lesssim$ 3]{s} to compute an energy spectrum of the extragalactic intensity from annihilation without halo substructure and $\epsilon=0.01$. Considering one level of substructure or choosing $\epsilon=10^{-3}$ doubles the running time.

\end{small}
\newpage
\setcounter{tocdepth}{2}


\section{Introduction \label{sec:intro}}

Indirect signatures of dark matter (DM) annihilation or decay are being actively sought for in the fluxes of charged cosmic rays \cite{2012CRPhy..13..740L}, \grs{} \cite{2016PhR...636....1C} and neutrinos \cite{2015JCAP...10..068T,Aartsen2017}.  The \clumpy{} code has been developed over the last decade to provide a public tool to estimate the DM-induced fluxes of \grs{} and neutrinos, in a large variety of objects and user-defined configurations. Our knowledge about the masses and shapes of DM structures throughout the Universe is largely dependent on cosmological simulations, which all provide similar but still different parametrisations of the DM properties; \clumpy{} allows the user to easily switch and combine these parametrisations and to efficiently compute resulting \gr{} or neutrino ($\nu$) signals for indirect DM searches. The goal of \clumpy{} is to eventually provide an unified and comprehensive calculation of the exotic signal at both the Galactic and extragalactic scales.

The first release of the code, published in \cite{2012CoPhC.183..656C},  focused on the optimised calculation of the so-called astrophysical \jfactor{}\footnote{Integration of the DM density squared (resp. density) along the line of sight for DM annihilation (resp. decay).}, a key quantity for indirect detection studies at the Galactic scale. The code could simply provide the \jfactor{} of a user-defined halo, or compute \jfactor{} maps in the flat sky approximation of the Galactic halo or of any isolated halo, including substructures. This first release was used in \cite{2011MNRAS.418.1526C,2011ApJ...733L..46W,2012PhRvD..85f3517C,2012MNRAS.425..477N,2012A&A...547A..16M}.

In \cite{2016CoPhC.200..336B}, a second and much more complete release of the code was delivered to the community. From then, \clumpy{} included the source \gr{} and $\nu$ spectra computed from \cite{2011JCAP...03..051C} that allow to compute not only the \jfactors{}, but the actual exotic prompt radiation from DM annihilation or decay. Another important new feature of that release was a Jeans analysis module to compute the DM profile of dwarf spheroidal galaxies, prime targets of indirect DM detection, from their kinematic data  and to propagate the uncertainties to the \jfactors{} (e.g., \cite{2015MNRAS.453..849B,2015ApJ...808L..36B,2015MNRAS.446.3002B,2016MNRAS.462..223B}). Using \healpix{}, the possibility to compute full skymaps was also included and largely used in  \cite{2016JCAP...09..047H}.

Up to now, the calculations implemented in \clumpy{} have been only valid at the Galactic scale and for the local Universe, where cosmology and absorption can be ignored.\footnote{At the Galactic scale, absorption can be safely ignored for photons below $\sim100$\;TeV, a regime compatible with the most studied DM candidates. However, this is not the case in PeV DM scenarios where the produced \grs{} will be significantly absorbed at the Galactic scale \cite{2015JCAP...10..014E}.\label{fn:galabsorb}} This paper describes the third  release of \clumpy{}, which now also includes the mean extragalactic contribution to the exotic \grs{} and neutrinos in a flat $\Lambda$CDM universe,  used to produce the results of \cite{2018JCAP...02..005H}. This addition is the missing piece to provide, within the same framework, a tool to consistently compare all potential DM targets on the Galactic and extragalactic scale.

The new extragalactic physics included into \clumpy{} is described in \secref{sec:calculation} of this paper. \secref{sec:implement} presents the main additions brought to the code in order to effectively implement this physics, while some new, more minor features included in this release are summarised in \secref{sec:new}. New parameters and keywords are described in \secref{sec:params}. \clumpy{} has several library dependencies that have made the previous release quite complex to install. For this third version of the code, the installation process has been greatly simplified, making some dependencies optional as described in \secref{sec:install}.

\section{Average $\gamma$-ray and $\nu$ fluxes from extragalactic DM\label{sec:calculation}}

The extragalactic differential \gr{} or $\nu$ intensity of annihilating Majorana DM\footnote{For annihilating Dirac-like particles, the factor $8\pi$ in the denominator of \eq{eq:mean_intensity_ann} is doubled to $16\pi$.}, averaged over the whole sky, is given by (see, e.g., \cite{2002PhRvD..66l3502U,2010JCAP...04..014A,2014PhRvD..89h3001N,2015JCAP...09..008T,2015PhR...598....1F,2016JCAP...08..069M})
\begin{eqnarray}
\label{eq:mean_intensity_ann} 
&I_{\rm ann}(E_{\gamma,\nu})\;=\;\left\langle\frac{\dd  \Phi}{\dd
E_{\gamma,\nu}\,\dd\Omega}\right\rangle_{\rm sky} 
\;= \;\frac{\overline{\varrho}^2_{\mathrm{DM},\,0} \,\sigmav}{8\pi\,m_{\chi}^2} 
\\
\times& \int\limits_0^{z_\mathrm{max}} c\,\dd z\,
\frac{(1+z)^3}{H(z)}\left\langle\delta^2(z)\right\rangle\left.\frac{\dd N^{{\gamma,\nu}}_{\mathrm{source}}}{\dd E_{\mathrm e}}\right|_{E_{\mathrm e}=(1+z)E_{\gamma,\nu}} \times e^{-\tau(z,\, E_{\gamma})}\,,\nonumber
\end{eqnarray}
with $E_{\gamma,\nu}$ the observed energy, $\Phi$ the flux, $\dd\Omega$ the elementary solid angle, $c$ the speed of light, and $\overline{\varrho}_{\mathrm{DM},\,0}$ the DM density of the Universe today. $H(z)$ is the Hubble constant at redshift $z$, $m_\chi$ the DM candidate mass, and $\sigmav$ the velocity averaged annihilation cross section. $\dd N^{\gamma,\nu}_{\mathrm{source}}/\dd E_{\mathrm e}$ is the differential \gr{}/$\nu$ yield per annihilation evaluated at $E_{\rm e}=(1+z)E_\gamma$, corresponding to a photon at $E_\gamma$ today, and $\tau(z,\,E_\gamma)$ describes the optical depth of the Universe to \grs{} of GeV to beyond-TeV energies. Finally, $\langle\delta^2 \rangle=1 + \mathrm{Var}(\delta)$ is the intensity multiplier related to the DM inhomogeneity $\delta$.\footnote{Density fluctuations in the  Universe, $\varrho_\mathrm{DM}(\Omega,\,z)=\delta(\Omega,\,z)\times \overline{\varrho}_\mathrm{DM}(z)$,  boost the rate of DM annihilations. For smoothly distributed DM, $\delta\equiv 1$, $\mathrm{Var}(\delta)=0$  and $\langle \delta^2 \rangle = 1$, whereas for a high density contrast, $\langle\delta^2 \rangle\approx \mathrm{Var}(\delta) \gg 1$.} 

For decaying DM with lifetime $\tau_\mathrm{DM}$, the intensity is given by 
\begin{eqnarray}
\label{eq:mean_intensity_dec} 
&I_{\rm decay}(E_{\gamma,\nu})
\;= \;\frac{\overline{\varrho}_{\mathrm{DM},\,0}}{4\pi\,\tau_\mathrm{DM}\,m_{\chi}} 
\\
\times& \int\limits_0^{z_\mathrm{max}} c\,\dd z\,
\frac{1}{H(z)}\,\left.\frac{\dd N^{{\gamma,\nu}}_{\mathrm{source}}}{\dd E_{\mathrm e}}\right|_{E_{\mathrm e}=(1+z)E_{\gamma,\nu}} \times e^{-\tau(z,\, E_{\gamma})}\,.\nonumber
\end{eqnarray}
As of now, \clumpy{} only accounts for the prompt \gr{} emission, given by the above equations. The diffuse secondary emission, produced by the inverse Compton scattering (ICS) of DM-produced $e^{+/-}$ with CMB photons or other radiation fields \cite{2009JCAP...07..020P,2009NuPhB.821..399C}, is not yet implemented. The spectrum of these up-scattered photons peaks at lower energies compared to the prompt $\gamma$-ray emission (see, e.g., figure~1 of \cite{2010PhRvD..81d3505B}); accounting for those photons would change the spectra produced by \clumpy{} at energies below 
$10^{-2} m_\chi$.\footnote{Similarly, in very heavy dark matter scenarios (10-100 PeV DM), a more involved process exists where prompt \gr{} emission produces energetic $e^+e^-$ pairs that will in turn generate \grs{} through ICS. Emission from these electromagnetic cascades is not included in this version of \clumpy{}, but may dominate the resulting \gr{} spectrum \cite{2018JCAP...04..060B}.}

\section{Implementation in \clumpy{}: new and updated functions \label{sec:implement}}

\subsection{New module {\tt cosmo.cc}}

Computation of the extragalactic intensity as given by \eq{eq:mean_intensity_ann} requires accounting for the cosmology. This is done in a new module, {\tt cosmo.cc}, which contains all cosmology-related functions, from the computation of the various cosmological distances to the evaluation of the halo mass function required by the intensity multiplier $\langle \delta^2(z) \rangle$, \eq{eq:delta_dm_annihil} below. A large fraction of the functions included in this module are a translation into {\tt C} of Eiichiro Komatsu's very useful Cosmology Routine Library in {\tt Fortran}.\footnote{\url{https://wwwmpa.mpa-garching.mpg.de/~komatsu/crl/}}

\subsubsection{Cosmological distances \label{subsubsec:distances}}
There are several ways to define distances in cosmology, whether one is interested in the i) distance between two objects along the line of sight (comoving distance), ii) the transverse distance between two objects at the same redshift (transverse comoving distance), iii) the physical size of an object given its observed angular size (angular diameter distance), or iv) the distance to link an observed flux to the intrinsic luminosity of a source (luminosity distance). These well-known definitions depend on the cosmological parameters and redshift and are not repeated here (see, e.g., \cite{1999astro.ph..5116H}), but have been included in the {\tt cosmo.cc}.

\subsubsection{Intensity multiplier  $\langle \delta^2(z) \rangle$ of annihilating DM}
The quantity $\langle \delta^2(z) \rangle$ is computed according to the halo model approach \cite{2002PhRvD..66l3502U,2013PhRvD..87l3539A,2016JCAP...08..069M}. In this setup, the intensity multiplier from the inhomogeneous Universe is written as
\begin{equation}
\left\langle\delta^2(z)\right\rangle = \frac{1}{\overline{\varrho}^2_{\mathrm{m,0}}}\; \int \dd M \frac{\dd n}{\dd M}(M,z)\times \mathcal{L}(M,z) \,,
\label{eq:delta_dm_annihil}
\end{equation}
where $\overline{\varrho}_{\mathrm{m,\,0}}$ is today's mean total matter density, ${\dd n}/\dd M$ the halo mass function, and $\mathcal{L}(M,z)$ the one-halo luminosity.

\paragraph{The halo mass function} The number density of haloes $ {\dd n}/{\dd M}$ in a given mass range at redshift $z$ depends on the variance $\sigma$ of the density fluctuations and on the multiplicity function $f(\sigma,\,z)$ that encodes nonlinear structure formation,
\beq
\frac{\mathrm{d}n}{\mathrm{d}M}(M,\,z)  = f(\sigma,z)\; \frac{\overline{\varrho}_{\rm m,0}}{M}\, \frac{\mathrm{d}\ln \sigma^{-1}}{\mathrm{d}M}\,.
\label{eq:halomassfunction}
\eeq
The variance $\sigma$ is calculated from the linear matter power spectrum, $P_{\rm lin}(k,z=0)$, according to
\beq
\sigma^2(M,\,z) = \frac{D(z)^2}{2\pi^2} \int P_{\rm lin}(k, z=0)\, \widehat{W}^2(kR)\,k^2\,\mathrm{d}k\,,
\label{eq:sigma2}
\eeq
with $R=[M/(\gamma_{\rm f}\,\overline{\varrho}_{\rm m,0})]^{1/3}$ the comoving scale radius of a collapsing sphere containing the mass $M$. Different shapes of the sphere can be selected (see \tab{tab:enum}), expressed in $k$-space by
\begin{align}
\widehat{W}(kR) &= 3\,(kR)^{-3} \,[\sin(kR)  - kR \,\cos(kR)],\;\gamma_{\rm f} = \frac{4\pi}{3} \label{eq:tophat}\tag{top-hat};\\
\widehat{W}(kR) &= \exp[- (kR)^2 / 2],\;\gamma_{\rm f} = (2\pi)^{3/2} \label{eq:gausswindow} \tag{Gaussian window};\\
\widehat{W}(kR) &= 1 - \theta(kR - 1),\;\gamma_{\rm f} = 6\pi^2 \label{eq:sharpk} \tag{sharp-$k$ filter}\,;
\end{align} 
with $\theta$ the Heaviside step-function. \clumpy{} is interfaced with the  \texttt{CLASS} code \cite{2011arXiv1104.2932L} to compute $P_{\rm lin}(k,z=0)$. The growth factor $D(z) = g(z)/g(z=0)$, defined from
\beq
g(z) = \frac{5}{2}\times \frac{8\pi G}{3}\overline{\varrho}_{\rm m,0}\times H(z)\,\int_z^\infty \frac{1+z'}{H^3(z')}\,\dd z'\,,
\label{eq:growth}
\eeq
then allows to compute the variance at any redshift ($G$ is the gravitational constant).

The multiplicity function $f(\sigma, z)$ is generally fitted to numerical simulations. \clumpy{} includes the parametrisations from a variety of simulations, considering only DM  or including baryons and relying on different cosmologies. The available parametrisations are listed in \tab{tab:enum} and the user may select any one of these with a corresponding keyword (see \secref{sec:params} for details).

\paragraph{The comoving one-halo luminosity}
The other term required by the intensity multiplier is the halo luminosity, defined to be
\beq
\mathcal{L}(M_{\Delta},z) = \int\dd V\, \rho^2_{\mathrm{halo}} = 4\pi \int_0^{R_{\Delta}}\dd r \,r^2\, \rho^2_{\mathrm{halo}}\,.
\label{eq:luminosity}
\eeq
It depends on the halo mass $M_{\Delta}$ (which defines its size $R_\Delta$, see \S\ref{subsec:delta}), on the halo density profile\footnote{The
radius $r_{-2}$ is defined to be $\dd \log \rho_{\mathrm{halo}}/\dd \log r|_{r=r_{-2}}=-2$, and $\rho_{-2} := \rho_{\mathrm{halo}}(r=r_{-2})$.} $\rho_{\mathrm{halo}}(r;\,\rho_{-2},\,r_{-2})$, and on the mass-concentration relation $c_{\Delta}(M_{\Delta},z) := R_{\Delta}/r_{-2}$ which is required to determine the normalisation of the profile given the halo mass. All mass-concentrations available in \clumpy{} are now implemented with their proper redshift-scaling. The halo luminosity has existed in \clumpy{} since the first release and is defined in the {\tt clumps.cc} module (see the previous release articles \cite{2012CoPhC.183..656C,2016CoPhC.200..336B}).

\subsubsection{Main functions}
\begin{itemize}
\item{\tt dh\_\{c, trans, a, l\}:} a set of functions returning, for a given cosmology and redshift, the comoving ({\tt c}), the transverse comoving ({\tt trans}), the angular diameter ({\tt a}) and the luminosity ({\tt l}) distances.

\item {\tt get\_pk:} for a given cosmology, loads the matter power spectrum from an existing file, the name of which should match a pattern obtained from the cosmological parameters. If the file does not exist, the function calls {\tt compute\_class\_pk} to compute $P_{\rm lin}(k)$ using {\tt CLASS} (which should have been previously installed by the user).

\item{\tt growth\_factor:} computes the linear growth rate $D(z)$, for a given redshift, using either the exact analytical solution of \eq{eq:growth} or the approximate (and faster to compute) formula of \cite{1992ARA&A..30..499C}, depending on the user's choice.

\item{\tt mf\_*:} a set of functions returning the multiplicity function obtained by various authors from various simulations, as listed in \tab{tab:enum}. The asterisk is to be replaced by the authors or simulation name.

\item {\tt compute\_massfunction:} returns a vector of tabulated values of the mass function (Eq.~\ref{eq:halomassfunction}, but written as $\mathrm{d}n/\mathrm{d}\ln M$) at a given redshift, for a given set of masses defined by the user in the main parameter file. The result for any mass is interpolated from these tabulated values.
\end{itemize}

\subsection{Absorption added in {\tt spectra.cc}\label{sec:absorption}}
Absorption of the \gr{} photons due to pair production after interaction with the extragalactic background light (EBL) and the cosmic microwave background (CMB)\footnote{Note that the implemented models do not consider  \gr{} absorption on the CMB, which affects the signal from nearby ($<\unit[1]{Mpc}$) sources above $\sim \unit[100]{TeV}$ (see \footref{fn:galabsorb}).} is another important term in \eq{eq:mean_intensity_ann}. Several types of models exist in the literature to describe the EBL. One approach, called backward modelling, starts from the observed multi-wavelength properties of existing local galaxies and extrapolate to larger redshifts (e.g.  \cite{2008A&A...487..837F, 2017A&A...603A..34F}). The so-called forward modelling relies on semi-analytical models of galaxy formation and evolution to compute the opacity at any redshift (e.g. \cite{2012MNRAS.422.3189G,2013ApJ...768..197I}). Other methods focus instead on the mechanisms responsible for the EBL, i.e. emission of stars and dust, and integrate over stellar properties and star formation rate \cite{2010ApJ...712..238F}, or rely on the direct observation of galaxies over the redshift range of interest \cite{2011MNRAS.410.2556D}. Given the variety in the modelling approaches, the results for $\tau(E,z)$ of each of the works cited above has been implemented in \clumpy{}. At each redshift, we transform the tabulated data from these works into coefficients of a fitted 11$^{\rm th}$ order polynomial in energy. The optical depth $\tau$ is then computed from the polynomial in energy and interpolated in redshift by the function {\tt opt\_depth} that has been added to the existing {\tt spectra.cc} module.
\subsection{New module {\tt extragal.cc}}

Relying on the two modules described in the last subsection, the {\tt extragal.cc} module puts everything together to compute the actual extragalactic exotic flux from \eq{eq:mean_intensity_ann} or (\ref{eq:mean_intensity_dec}). As seen from the equations above, this requires several nested integrals that would be very time consuming to compute if done naively. To optimise this integration, \clumpy{} first tabulates the mass function and other cosmological quantities on a grid of masses and redshifts and then performs interpolations to obtain the results for any $(M,z)$ combination during integration. By default, the grid is calculated on 1000 mass nodes and $\Delta z = 0.1$ within a couple of seconds; to further increase speed or precision, the resolution of the grid can be adjusted by the user in the main parameter file.

\subsubsection{Main functions}

\begin{itemize}
\item {\tt init\_extragal:} this function initialises the computation of the extragalactic flux by tabulating the values of the various distance definitions (\secref{subsubsec:distances}) on the redshift grid and the values of the mass function on the 2D $(M,z)$ grid. The resulting arrays are stored as global variables.

\item {\tt d\_intensitymultiplier\_dlnM:} this function uses the previously tabulated values to interpolate the mass function to the relevant mass and redshift and returns the corresponding intensity multiplier, \eq{eq:delta_dm_annihil}, but under the form $\dd\langle\delta^2\rangle/\dd\ln M$.

\item {\tt dPhidOmegadE:} this returns the result for the extragalactic differential \gr{} intensity of DM annihilation, according to \eq{eq:mean_intensity_ann}.

\item {\tt analyse\_*:} a series of functions called by the new {\it extragalactic submodules} ({\tt -e}) available from the user interface menu as presented in \secref{sec:install}. These allow quick checking and plotting of various intermediate and final quantities, such as cosmological distances, the intensity multiplier, EBL absorption, or differential flux.

\end{itemize}

\section{Additional new features \label{sec:new}}

\subsection{Signal from individual  high-redshift extragalactic haloes}

Thanks to the implementation of cosmological distances and geometry, \clumpy{} now also allows the rigorous computation of the fluxes from individual haloes  located at $z>0$. Assuming the haloes span is small enough ($\Delta z/z\ll 1$), one may separate spectral and astrophysical contributions in the flux calculation. For annihilating Majorana\footnote{For Dirac-DM, the factor $8\pi$ in the denominator of \eq{eq:flux_1halo_ann}  is $16\pi$.} DM, the flux from an individual halo reads 
\begin{eqnarray}
\label{eq:flux_1halo_ann} 
\frac{\Phi_{\rm ann}}{\dd E_{\gamma,\nu}}(E_{\gamma,\nu},\,\Delta\Omega)
\;=& \frac{\sigmav}{8\pi\,m_{\chi}^2} \, \left.\frac{\dd N^{{\gamma,\nu}}_{\mathrm{source}}}{\dd E_{\mathrm e}}\right|_{E_{\mathrm e}=(1+z)E_{\gamma,\nu}} \times e^{-\tau(z,\, E_{\gamma})}
\nonumber
\\ 
\times& (1+z)^3\; \int_0^{\Delta \Omega} \int_{\rm l.o.s.} \,\rho_{\rm halo}^2\,\dd l \,\dd \Omega
\,,
\end{eqnarray}
and similarly, the flux from decaying DM is
\begin{align}
\label{eq:flux_1halo_dec} 
\frac{\Phi_{\rm decay}}{\dd E_{\gamma,\nu}}(E_{\gamma,\nu},\,\Delta\Omega)
\;=& \;\frac{1}{4\pi\,\tau_\mathrm{DM}\,m_{\chi}} \, \left.\frac{\dd N^{{\gamma,\nu}}_{\mathrm{source}}}{\dd E_{\mathrm e}}\right|_{E_{\mathrm e}=(1+z)E_{\gamma,\nu}} 
\times e^{-\tau(z,\, E_{\gamma})} \nonumber
\\
\times&\; \int_0^{\Delta \Omega} \int_{\rm l.o.s.} \,\rho_{\rm halo}\,\dd l \,\dd \Omega
\,.
\end{align}
In  \eqs{eq:flux_1halo_ann}{eq:flux_1halo_dec}, the line-of-sight distance (l.o.s.) $l$ to the object now corresponds to the comoving distance, while $\rho_{\rm halo}(r)$ is given in comoving coordinates. From the second line in each equation, one may define $J/D$-factors analogously to that computed for the local Universe by the previous versions of the code.

\subsection{Choice of various overdensity definitions $\Delta$}
\label{subsec:delta}

In the previous versions of \clumpy{}, the halo radii $R_{\Delta}$ and masses $M_{\Delta}$ were defined according to the findings of \cite{1998ApJ...495...80B} (`virial' radii and masses). 
In general, the mass $M_{\Delta}$ of a halo is connected to its size, $R_{\Delta}$, via the relation
\beq
R_{\Delta}( M_{\Delta}, z) =\left( \frac{3\, M_{\Delta}}{ 4\pi \times \Delta(z) \times \varrho_{\rm c}(z)}\right)^{1/ 3} \times (1 + z)\,,
\label{eq:rdeltamdelta}
\eeq
where the subscript $\Delta$ denotes a characteristic collapse overdensity above the critical density of the Universe, $\varrho_{\rm c}= 3H^2(z)/8\pi G$. In the new \clumpy{} release, all calculations can be chosen to be performed with respect to any of the following overdensity definitions:
\begin{align}
\Delta(z) & = {\rm const.}  \label{eq:deltacrit},\\
\Delta(z) &= {{\rm const.} \times \Omega_{\rm m}(z)} =: {\Delta_{\rm m} \times \Omega_{\rm m}(z)}\label{eq:deltam},\\
\Delta(z) &= 18\pi^2 + 82\,[\Omega_{\rm m}(z) - 1)]- 39\,[\Omega_{\rm m}(z) -
1]^2 \label{eq:bryannorman}\,.
\end{align} 
Note that \eq{eq:bryannorman} corresponds to the previously only available description from \cite{1998ApJ...495...80B} (valid in a flat Universe). If a mass-concentration $c_{\Delta}(M_{\Delta},z) $  or halo mass function ${\mathrm{d}n}/{\mathrm{d}M}(M,\,z) $ are natively provided for a $\Delta$ different from the user's choice, $c_{\Delta}(M_{\Delta},z) $ or ${\mathrm{d}n}/{\mathrm{d}M}(M,\,z) $ are internally rescaled to the user-chosen $\Delta$. This rescaling depends on the user-defined halo profile and mass-concentration relation; we refer the reader to appendix~A of \cite{2018JCAP...02..005H} for the corresponding equations and algorithm.

\subsection{Different  scale radii of the substructure distribution and  host halo density}
\label{sec:dpdvscaleradius}

In the previous versions, the scale radius of the spatial distribution of subhaloes in the host, $\dd \mathcal{P}_V/\dd V = \rhosub/M_{\rm subs}$, $M_{\rm subs}=f_{\rm subs}\,M_{\rm halo}$ \cite{2012CoPhC.183..656C,2016CoPhC.200..336B},  was not an available input parameter. Instead, it was internally set to the scale radius of the total DM density distribution, $\rhotot$. In the new release of the code, the input parameter \texttt{gTYPE\_SUBS\_DPDV\_RSCALE\_TO\_RS\_HOST} (with \texttt{TYPE = DSPH, MW, GALAXY, CLUSTER, EXTRAGAL}) is available to freely choose $r_{\rm s,\,\rhosub} = a\,r_{\rm s,\,\rhotot}$ as a fraction or multiple $a$ of the host halo's scale radius. Note that when studying a population of sub-subhaloes in single objects (\texttt{DSPH, MW, GALAXY, CLUSTER}) or halo substructure for the mean extragalactic signal  (\texttt{EXTRAGAL}), no mass or redshift dependence of the ratio $a$ is implemented.

\subsection{New $J$-factor normalisation of 2D skymaps}

Since its first release, \clumpy{} includes a module which allows the computation of two-dimensional (2D) skymaps of the $J$-factor. While this module has been extended in \cite{2016CoPhC.200..336B} for large-sky patches, the 2D pixel values in the maps were always given with respect to a user-defined integration region, which did not coincide with the pixel size. This format has now changed: via the \texttt{gSIM\_HEALPIX\_NSIDE} input parameter, a map resolution is chosen, namely $\Delta \Omega = \unit[\pi/3\,\nside^{-2}]{sr}$ (constant in size for all pixels), and we provide $J(\Delta \Omega)$ or the flux $\Phi(\Delta \Omega)$ in each pixel.\footnote{We continue to use an oversampled box-smoothing of the $J$-factor in 2D-maps, as detailed in the technical documentation of the code.} Note that we continue to additionally provide to the user $\dd J/\dd \Omega$ or $\dd \Phi/\dd \Omega$ in the 2D output \fits{} files.

\subsection{Conversion of the 2D map outputs from \healpix{} pixelisation to \fits{} images} 

\begin{figure}[!t]
\centering
\includegraphics[width=0.58\columnwidth]{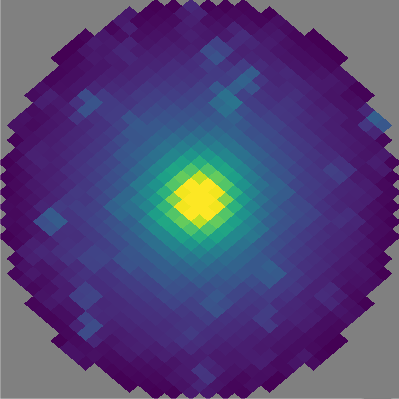}\vspace{0.1cm}
\includegraphics[width=\columnwidth]{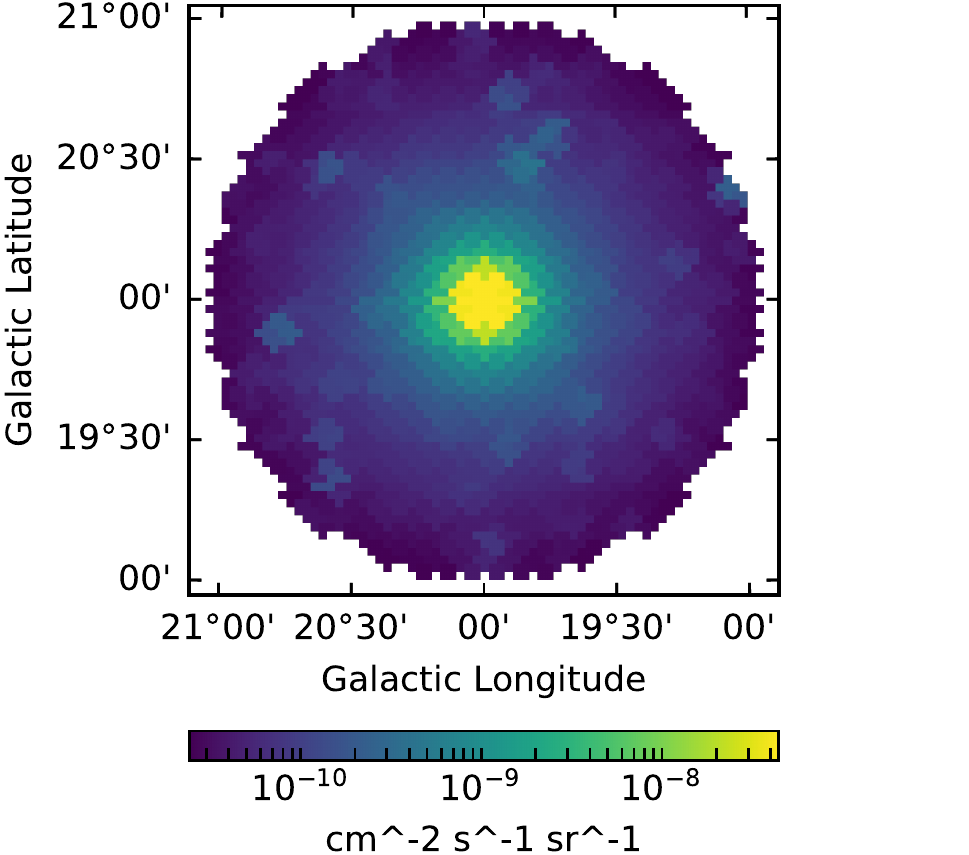}
\caption{Illustration of the 2D map transformation using the {\tt makeFitsImage.py} script. {\em Top panel:}  2D map output in the \healpix{} format. Shown is the integrated \gr{} intensity of an example halo, drawn for the standard parameters in the {\tt -h5 module}, using {\tt matplotlib} and the {\tt healpy.cartview()} function. {\em Bottom panel:} Transformed \fits{} image, displayed with a simple \python{} script using {\tt matplotlib} and {\tt astropy}. Examples to generate above figures can be found in the \clumpy{} online documentation.}
\label{fig:example_2D}
\end{figure}

Since the last code release \cite{2016CoPhC.200..336B}, 2D maps are stored in the \fits{} format and \healpix{} pixelisation; these maps can be read and displayed by common programs handling the  \fits{}/\healpix{} formats. To further ease the post-processing and plotting of the 2D maps, we now provide with the code a \python{} script to convert \healpix{} maps into rectangular \fits{} images in Cartesian projection. This {\tt makeFitsImage.py} script relies on the {\tt healpy}\footnote{\url{http://github.com/healpy/healpy/}} (for the projection and regridding) and {\tt astropy}\footnote{\url{http://www.astropy.org/}} (for the \fits{} I/O) packages. The transformed images can also be directly read by common \fits{} viewers and analysis packages as, e.g., {\tt gammalib} \cite{2016A&A...593A...1K}. However, the transformation from \healpix{} to \fits{} images is non-reversible and degrading, also, the property of equi-areal pixels is lost. This is illustrated in \fig{fig:example_2D}. For the extended \gr{} emission of an arbitrary example halo, we present the original \healpix{} output  at the top and the projected image at the bottom. It can be seen how an oversampled, rectangular grid is rastered in the projected image. Note that a coarse resolution ($\nside=1024$) is chosen in these figures for illustration purpose. For an increased resolution, the visible difference between the figures vanishes and the radial symmetry of the object is more pronounced.

\section{New parameters and keywords}
\label{sec:params}

As in previous releases, \clumpy{} comes with a large number of parameters required to perform any run: these include physics constants, dark matter halo properties,  and simulation configurations. Some parameters require keywords that allow the user to select preset ingredients  from the literature.

\input{table_params}
\paragraph{Parameters}

In this release, we have made the following changes for the list of parameters, as shown in \tab{tab:params}:
\begin{itemize}
  \item For a better consistency and readability, we renamed some existing parameters (Milky way and generic halo structural parameters);

  \item In the course  of homogenising the input parameter control as described in \secref{sec:refactoring}, many arguments previously solely specified via the command line are now given explicit parameter names.

  \item The most important changes are those associated with the many new parameters required to perform the extragalactic intensity calculation described in \secref{sec:calculation}. We now have a more complete description of the cosmological parameters, extragalactic haloes structural parameters, and simulation parameters for the extragalactic calculation (lower part of the table).
\end{itemize}

\input{table_enums}
\paragraph{Keywords} The changes for the lists of keywords are reported in \tab{tab:enum}:
\begin{itemize}
  \item The upper half of the table gathers the keywords associated to a model or parametrisation for the new extragalactic ingredients discussed in \secref{sec:implement}. They correspond to the absorption model (\secref{sec:absorption}), the model to describe $\Delta(z)$ \eqpar{eq:deltacrit}, the growth factor \eqpar{eq:growth}, the halo mass function parametrisation \eqpar{eq:halomassfunction}, and the window function appearing in \eq{eq:sigma2}.
  \item The lower half of the table repeats the keywords to enumerators already presented in the previous version, with names having been changed for consistency (in particular for subhalo spatial distributions) and new keywords/parametrisations having been added. These changes are highlighted in grey.
\end{itemize}

\section{Installation and code execution \label{sec:install}}

We highlight in this section the changes made in this new release. Detailed instructions about the code structure, and how to install and execute it, are provided in the online documentation at \url{http://lpsc.in2p3.fr/clumpy/}.

\subsection{Code installation, tests, and documentation\label{sec:code}}

Several improvements have been made for the download, installation, and use of the \clumpy{} C/C++ code.
\begin{itemize}
\item Public version control: the code is now under \git{}\footnote{\url{http://git-scm.com/}} version control, publicly accessible and  currently  at \url{https://gitlab.com/clumpy/CLUMPY}, from where the latest development versions can be directly retrieved. 
\item Compilation and dependencies: this new release of \clumpy{} (C/C++) is compiled with {\tt cmake}\footnote{\url{http://cmake.org/}}. It now only depends on the \cfitsio{} and \gsl{} libraries, with the \rootcern{} library optional (some sub-modules, pop-up graphics, and outputs in \rootcern{} format are available only when \rootcern{} is installed). The dependency on \healpix{} \cite{2005ApJ...622..759G} is now ensured by a frozen version of the library (version 3.31) shipped with the code, which is internally built at compilation.
\item General and code documentation: we have completely revisited the documentation, now based on {\sc sphinx}\footnote{\url{www.sphinx-doc.org}}. For developers, \doxygen{}\footnote{\url{http://www.stack.nl/~dimitri/doxygen/}} pages  are additionally generated and available from the documentation pages.
\item Integration tests: we now provide an automated test suite, {\tt ./bin/clumpy\_tests}, to check the proper output of all modules after the installation.
\end{itemize}

\subsection{Code structure and executables}

As in the previous versions, declarations are stored in {\tt include/*.h} (with the two new libraries {\tt cosmo.h} and {\tt extragal.h}), sources in {\tt src/*.cc}, compiled libraries are saved in {\tt lib/}, and executables in {\tt bin/}. Tabulated tables on which \clumpy{} relies (EBL, $P(k)$, PPPC4DMID spectra \cite{2014JCAP...03..053B}, reference test files\dots) are located in {\tt data/}.

Beside the two executables for the Jeans analysis \cite{2016CoPhC.200..336B} ({\tt bin/clumpy\_jeansChi2} and {\tt bin/clumpy\_jeansMCMC}) and the new module to test the successful installation ({\tt bin/clumpy\_tests}), the main executable is {\tt bin/clumpy}, whose submodules are reproduced below for completeness: 
   \begin{itemize}
     \item {\tt ./bin/clumpy -g[index]}: galactic calculations,
     \item {\tt ./bin/clumpy -h[index]}: isolated/list of haloes,
     \item {\tt ./bin/clumpy -s[index]}: run on `statistical' files,
     \item {\tt ./bin/clumpy -o[index]}: skymap file manipulation,
     \item {\tt ./bin/clumpy -z}: $\gamma$ and $\nu$ spectra,
     \item {\tt ./bin/clumpy -f}: convert $J$-factor skymap to flux map,
   \end{itemize}
and a new submodule provided in this release:
   \begin{itemize}
     \item {\tt ./bin/clumpy -e[index]}: triggers the new extragalactic module and prints a list of available submodules ({\tt e0} to {\tt e6}) if no index is put.

     \item {\tt ./bin/clumpy -e[index] -D}: generates a file {\tt clumpy\_params\_e[index].txt} containing all parameters and default values to execute the corresponding run. This is particularly useful for new users to familiarise themselves with the required ingredients necessary to each requested computation.
   \end{itemize}

Running all executables without further options results in self-explanatory messages with instructions of how to run the various modules. Some explicit examples are given in \secref{subsec:examples}.

\subsection{Upgraded input parameter interface \label{sec:refactoring}}

To allow a better control of the manual execution of \clumpy{} and a better interface to a pipelined execution of the code, \clumpy{} input parameters can now be fully and flexibly controlled via the command line.

All parameters required for a specific run can be set in a parameter file, following the same format since the first version of the code\footnote{The backwards-compatibility of parameter files from the previous releases is still lost due to new parameters required replacing the former command-line syntax, as well as deprecated and renamed parameters as described in \secref{sec:params}.} or can alternatively be parsed via the command line; specifying a parameter value in the command line is always preferred over the parameter file (with a warning being drawn in the case of a double parameter setting). Also, the code now only requires the parameters needed for the specific executed run (and a warning is drawn for set, but unrelated parameters). In some cases, the requirement of a parameter depends on the value of another parameter, e.g., the number of shape parameters for a given DM density profile. These dependencies are now checked by the \texttt{load\_parameters()} function at the beginning of each run execution, and default values are proposed after abort due to missing parameters. Also, default parameter files can be generated individually for each simulation mode with the \texttt{-D} flag.

This comprehensive expansion of the programming interface will finally also ease the construction of a wrapping \python{} module planned for a future release.

\subsection{Run examples: extragalactic module\label{subsec:examples}}

\begin{figure}[!t]
\centering
\includegraphics[width=\columnwidth]{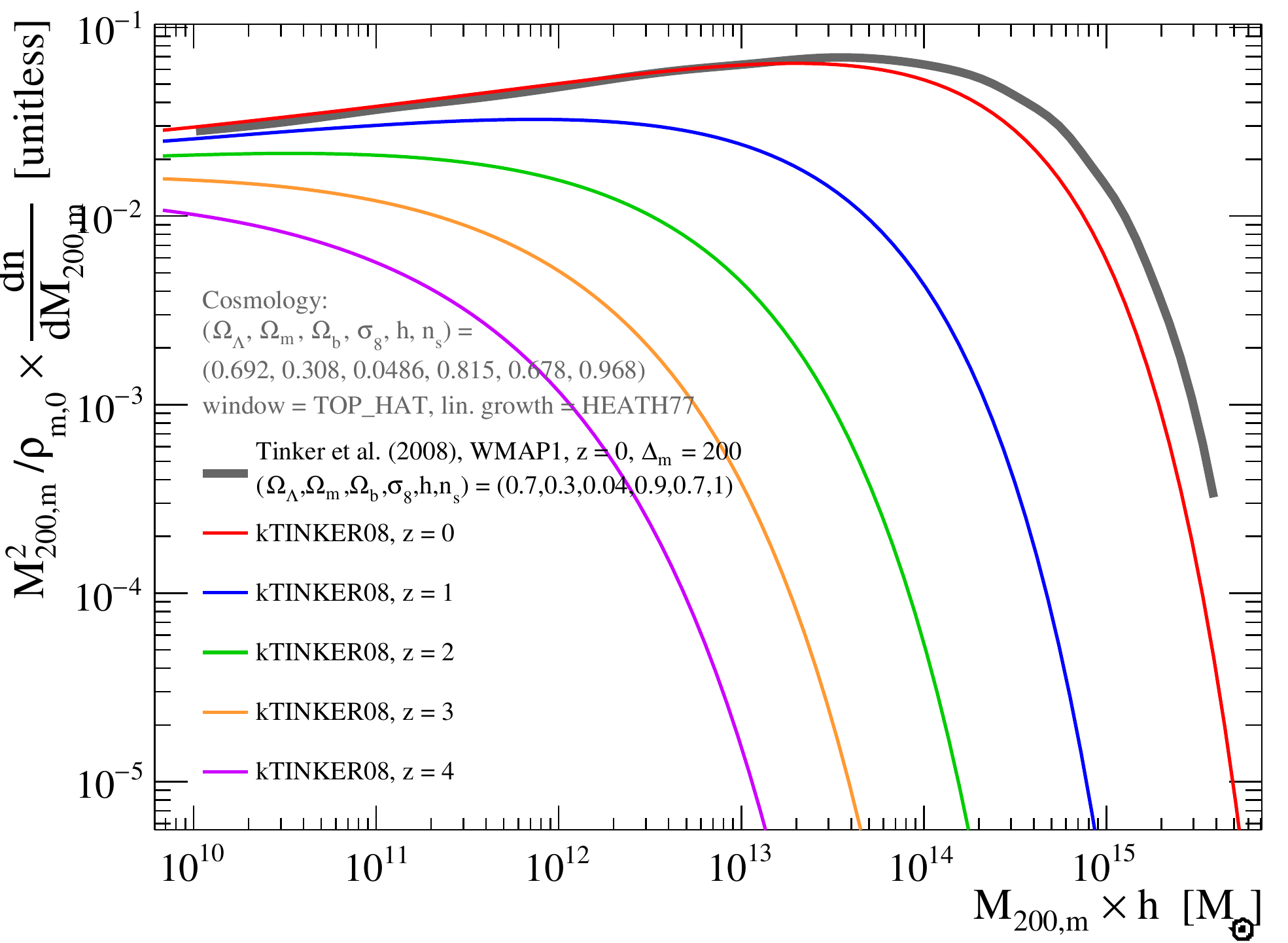}
\includegraphics[width=\columnwidth]{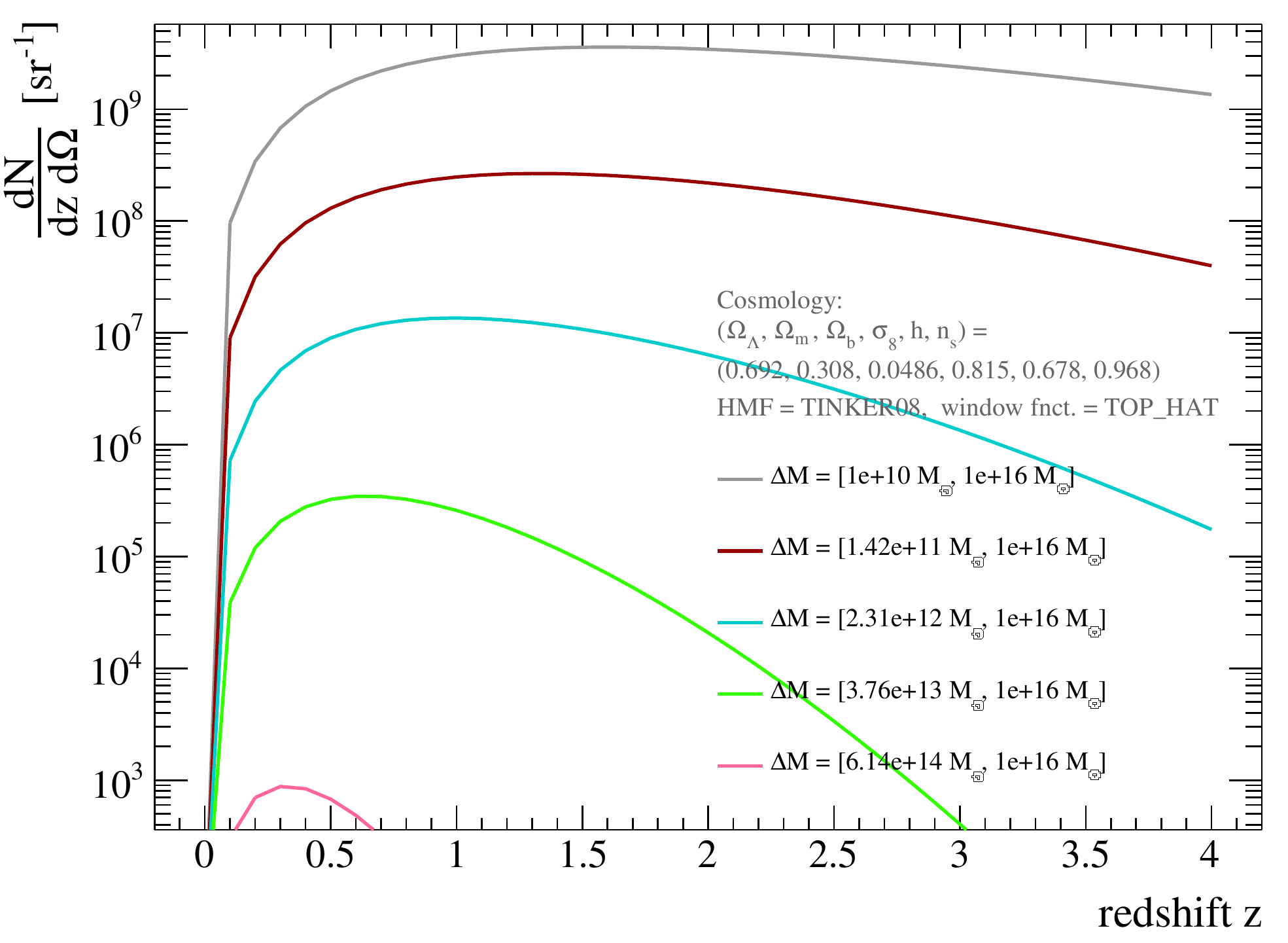}
\caption{Pop-up graphics from the extragalactic {\tt -e1} module to study the halo mass function, \eq{eq:halomassfunction}. {\em Top panel:} Halo mass function from \cite{2008ApJ...688..709T} ({\tt kTINKER08}) at various redshifts $z$ between $z=0$ and $z=4$ and \planck{} 2015 \cite{2016A&A...594A..13P} cosmological parameters. For comparison, the mass function at $z=0$ from the original publication \cite{2008ApJ...688..709T} (and the cosmology chosen therein) is overplotted in grey. {\em Bottom panel:} Halo occupation (average number of haloes per redshift shell and solid angle) for different mass ranges, derived from the above halo mass function. In these plots,  halo masses are defined with respect to the mean density, $\Delta(z) = 200\times \Omega_{\rm m}(z)$, see \eq{eq:deltam}.}
\label{fig:example_e1}
\end{figure}

\begin{figure}[!t]
\centering
\includegraphics[width=\columnwidth]{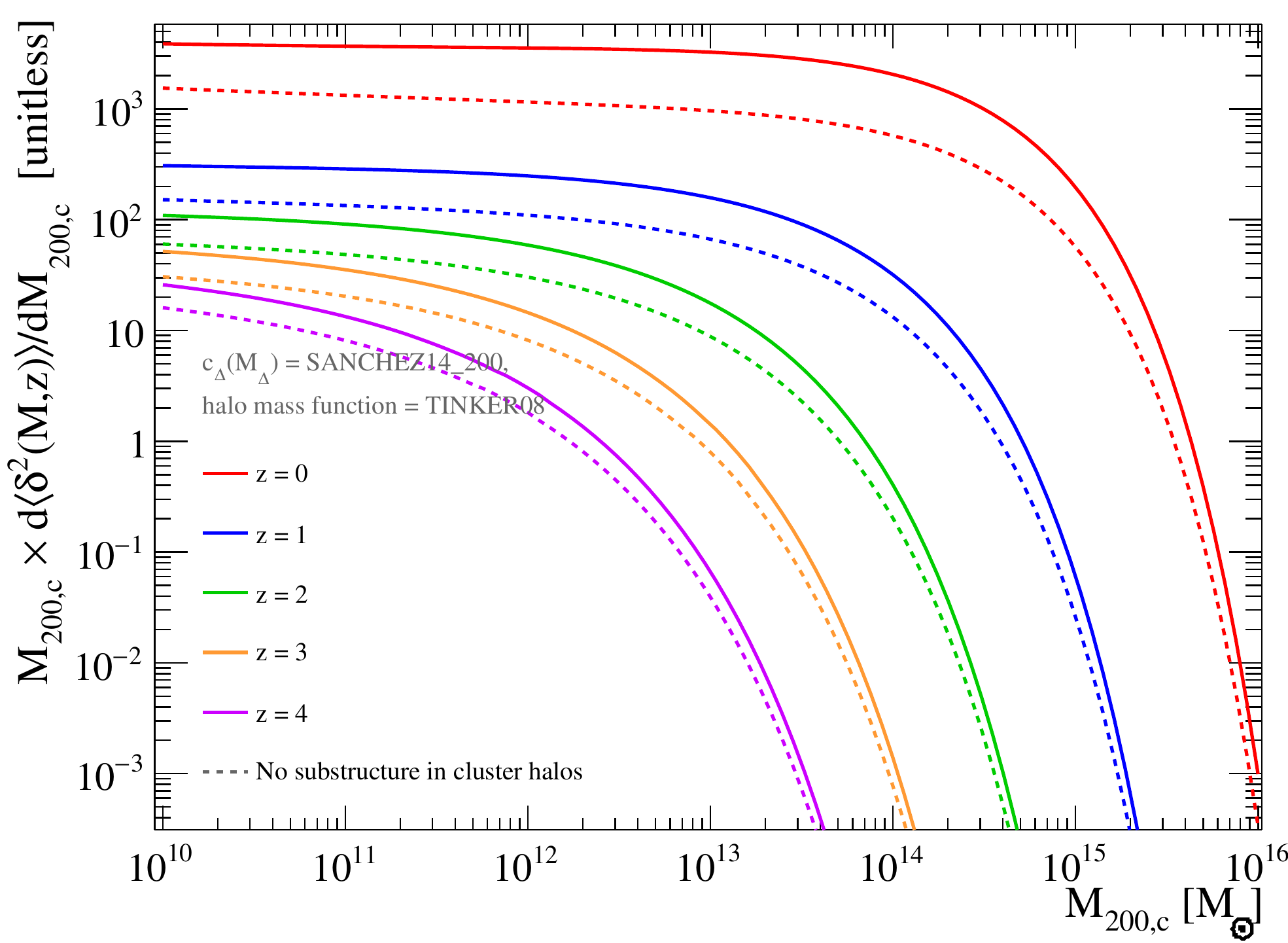}
\caption{Pop-up graphic from the extragalactic {\tt -e3} module to study the intensity multiplier, \eq{eq:delta_dm_annihil}, at various redshifts $z$ between $z=0$ and $z=4$. The solid lines show the intensity multiplier with the boost from halo substructure, while the dashed lines assume zero substructure boost. This result relies on the {\tt kTINKER08} \cite{2008ApJ...688..709T} mass function and {\tt kSANCHEZ14\_200} \cite{2014MNRAS.442.2271S} $c_{\Delta}(M_{\Delta},z)$ parametrisation. The intensity multipliers are plotted over the halo masses with respect to the critical density, $\Delta(z) = 200$, see \eq{eq:deltacrit}.}
\label{fig:example_e3}
\end{figure}

\begin{figure}[!t]
\centering
\includegraphics[width=\columnwidth]{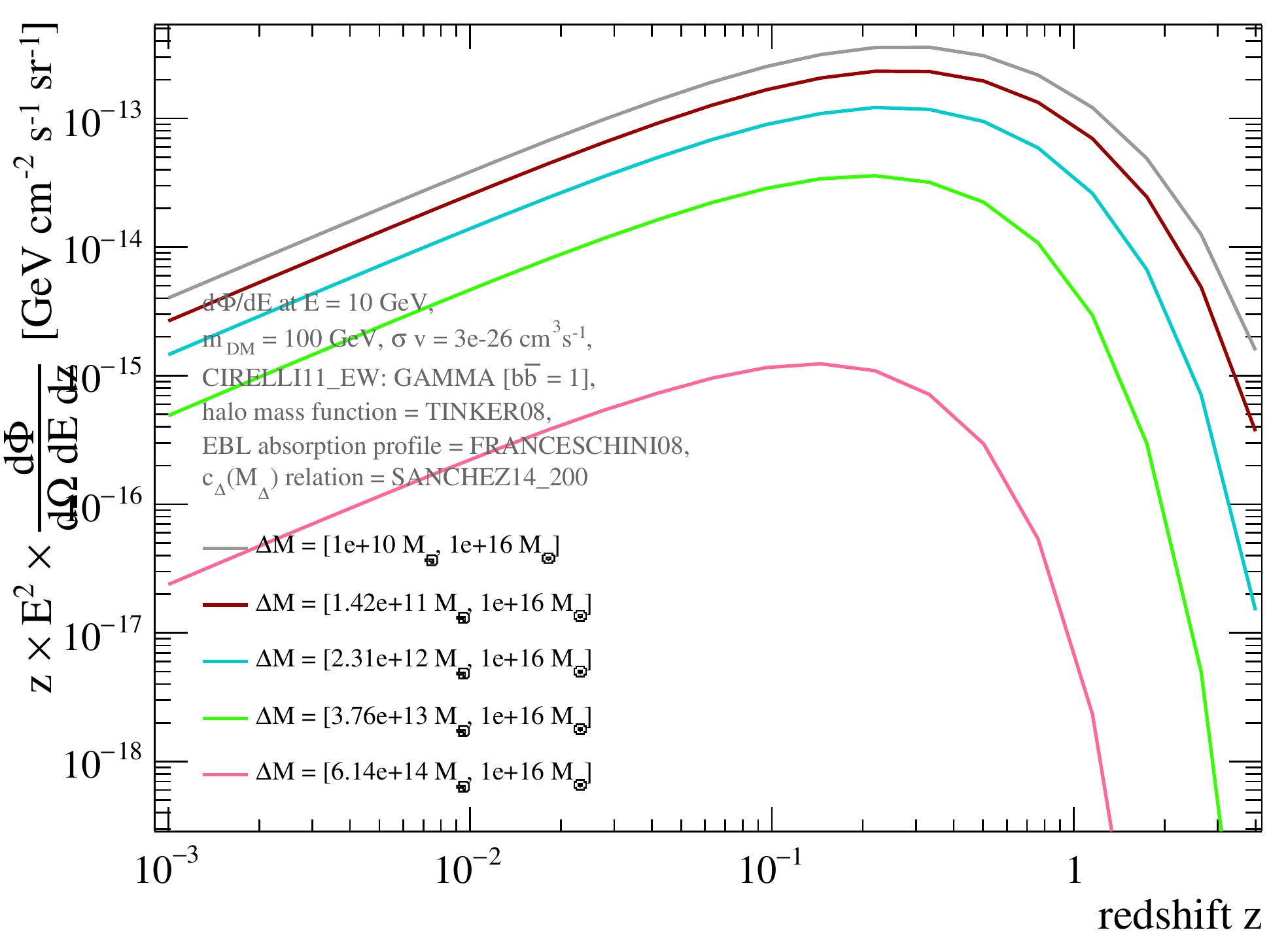}
\caption{Pop-up graphic from the extragalactic {\tt -e5} module to study different redshift- and mass-range contributions to the \gr{} intensity  $I =\dd \Phi/\dd E/\dd\Omega$.}
\label{fig:example_e5}
\end{figure}

\begin{figure}[!t]
\centering
\includegraphics[width=\columnwidth]{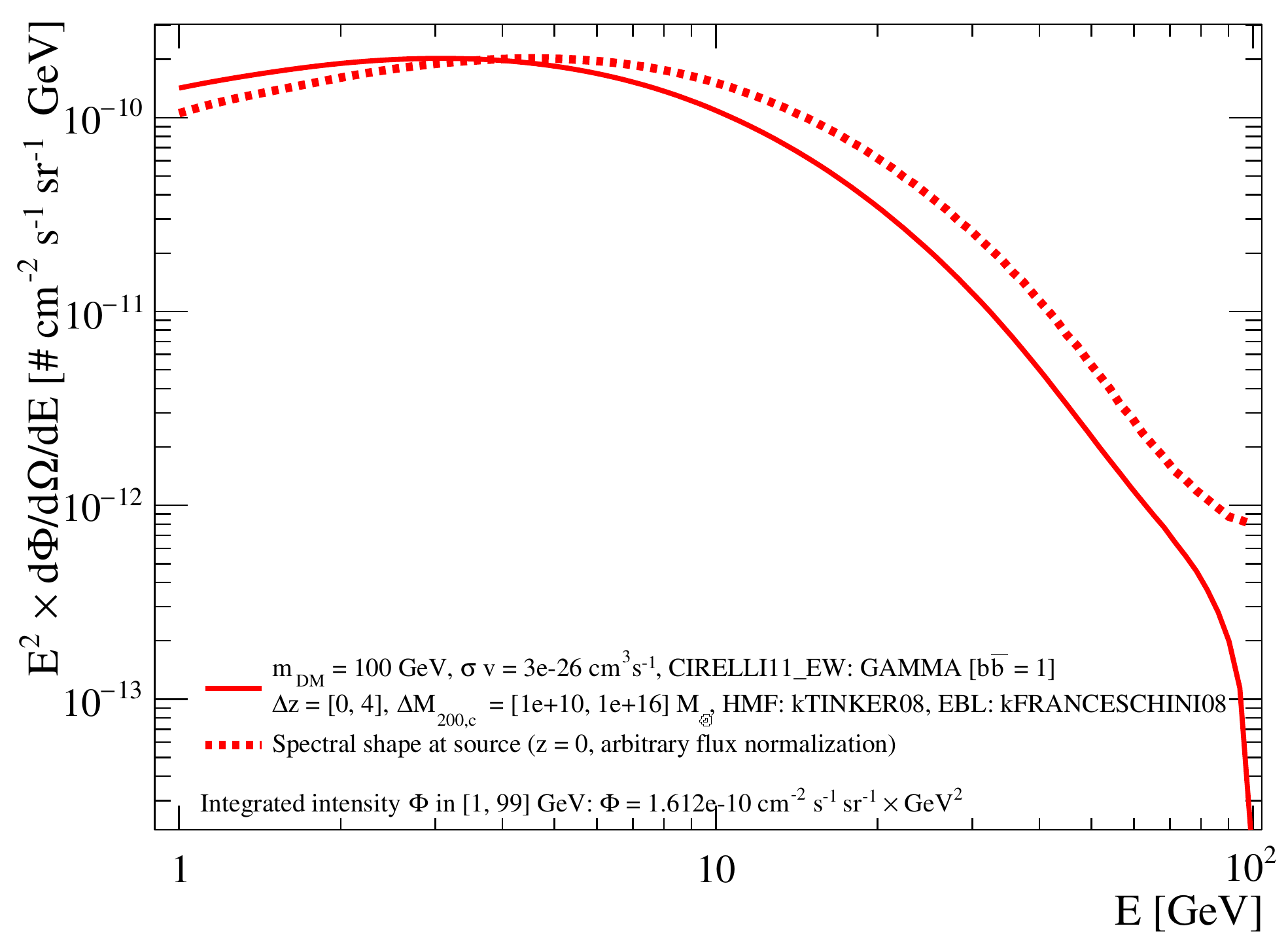}
\caption{Pop-up graphic with the result of the extragalactic {\tt -e6} module and standard parameters to obtain the mean \gr{} intensity from extragalactic DM annihilation, \eq{eq:mean_intensity_ann} (solid line). The dashed line shows the spectral shape for the chosen annihilation channel ($\chi\chi\rightarrow b\bar{b}$) at $z=0$ and arbitrary normalisation. Note that in line with the user-specified multiplication of the differential intensity with $E^n$, the integrated intensity reported in the figure is calculated as $\Phi=\int E^n\,\dd \Phi/(\dd\Omega\,\dd E)\,\dd E$ (e.g., for $n=1$, $\Phi$ corresponds to the energy flux).}
\label{fig:example_e6}
\end{figure}

To illustrate the execution of the new extragalactic module and the refactored input interface described in the last paragraph, we present some examples below, for various indices for the submodules. For all modules,  if \clumpy{} is linked to  \rootcern{}, interactive pop-up graphics are directly displayed and/or the results are saved in \rootcern{} format (results are always written to {\tt ASCII} files). If not otherwise stated, below examples with default parameters provide results and pop-up graphics (like the shown \figs{fig:example_e1}{fig:example_e6}) on the fly within a few seconds.
   \begin{itemize}
     \item {\tt ./bin/clumpy -e0 -i clumpy\_params\_e0.txt}: computes the cosmological comoving, angular diameter, and luminosity distances, as well as the $\Omega$ and Hubble parameters at $z>0$,  with all necessary input parameters read from {\tt clumpy\_params\_e0.txt}.

     \item {\tt ./bin/clumpy -e1 -i clumpy\_params\_e1.txt}: computes the halo mass function, \eq{eq:halomassfunction}. Two of the resulting \rootcern{} pop-up graphics are displayed in \fig{fig:example_e1}.  Note that a file\footnote{\tt Y\_hX\_OmegaBX\_OmegaMX\_OmegaLX\_nsX\_tauX\_z0\_lin.dat} is needed in the directory {\tt data/pk\_precomp/} containing $P_{\rm lin}(k,z=0)$, where {\tt X} denote the cosmological parameters specified in {\tt clumpy\_params\_e1.txt} and {\tt Y} is an arbitrary prefix. If no file is found, \clumpy{} tries to run \texttt{CLASS} to generate it (with {\tt Y $=$ class}). Any parameters in the parameter file can be overridden from the command line\footnote{For instance, {\tt ./bin/clumpy -e1 -i clumpy\_params\_e1.txt --gSIM\_EXTRAGAL\_ZMAX=8} overrides {\tt gSIM\_EXTRAGAL\_ZMAX} with the value 8 (default is 4), calculating the mass function up to $z_{\rm max}=8$.}.

     \item {\tt ./bin/clumpy -e2 -i clumpy\_params\_e2.txt}: Computes $c_{\Delta}(M_{\Delta},z)$, the luminosity $\mathcal{L}(M_{\Delta},z)$ \eqpar{eq:luminosity}, and the boost\footnote{The boost is computed self-consistently once the user has specified all relevant substructure quantities (substructure halo profile, mass distribution, spatial distribution, mass concentration relation). No pre-defined boost parametrisation under the form $\mathcal{B}(M_\Delta,z)$, as sometimes found in the literature, is implemented in this release.}, $\mathcal{L}_{\rm halo\,substructure}/\mathcal{L}_{\rm no\,substructure}$, over a user-defined mass range of extragalactic haloes. With this submodule, Fig.~1 from \cite{2016CoPhC.200..336B} is reproduced with a run time of about two minutes.

     \item {\tt ./bin/clumpy -e3 -i clumpy\_params\_e3.txt}: computes the intensity multiplier, \eq{eq:delta_dm_annihil}. The \rootcern{} pop-up display with the result for the default parameters is shown in \fig{fig:example_e3}.

    \item {\tt ./bin/clumpy -e4 -i clumpy\_params\_e4.txt}: computes the \gr{} EBL absorption exponent, $\tau(z,\,E_\gamma)$, and $e^{-\tau(z,\,E_\gamma)}$ from tabulated values (see \secref{sec:absorption}).

    \item {\tt ./bin/clumpy -e5 -i clumpy\_params\_e5.txt}: computes the differential contribution to the \gr{} or $\nu$ intensity from extragalactic DM, \eq{eq:mean_intensity_ann} or (\ref{eq:mean_intensity_dec}), from different redshift and mass shells, $\dd I/\dd z(z;\, \Delta M,\,E_{\gamma,\nu})$, and $\dd I/\dd M(M;\, \Delta z,\,E_{\gamma,\nu})$. This allows to get a better grasp of the contribution of different redshift and halo mass decades to the radiation intensity. For example, \fig{fig:example_e5} shows that at $E_{\gamma} = \unit[10]{GeV}$, for the chosen branching channel and the displayed mass ranges, the `brightest' redshift shell contributing to the \gr{} intensity lies between $0.2\lesssim z\lesssim 0.5$.

    \item {\tt ./bin/clumpy -e6 -i clumpy\_params\_e6.txt}: fully integrates the \gr{} or $\nu$ intensity $I_{\rm ann}(E_{\gamma,\nu})$ from extragalactic DM, \eq{eq:mean_intensity_ann} or \eq{eq:mean_intensity_dec}. The  result for the default parameters, annihilation according to \eq{eq:mean_intensity_ann}, is shown in \fig{fig:example_e6}.
   \end{itemize}

We have checked to the best of our ability that the \clumpy{} calculations match the results previously found in the literature. For example, the red curve in the top panel of \fig{fig:example_e1} does match the grey reference curve from \cite{2008ApJ...688..709T} when using the appropriate cosmology. Similarly, in Fig.~6 of \cite{2018JCAP...02..005H}, we compared the extragalactic fluxes computed with this version of \clumpy{} to several existing results and discuss the agreement or discrepancies in terms of parameter choices.
\section{Conclusions \label{sec:conclusion}}

After ten years of development, the \clumpy{} code now provides a comprehensive framework to compute indirect \gr{} and $\nu$ signals (from DM annihilation or decay) from the Galactic to the extragalactic scales.

In this new release, the code allows the flux and intensity calculation from haloes at any redshift, and to compute the average diffuse emission from all extragalactic DM. In doing so, the computation of various cosmological quantities had to be implemented, like halo mass functions with respect to any overdensity definition $\Delta$ and the optical depth to \grs{} due to pair production at the EBL. These intermediate quantities can be computed on their own, without seeking for \gr{} or $\nu$ signals from DM. By providing the corresponding submodules, we hope to widen the usage of the code at the interface between the astrophysics, astroparticle, and cosmology communities.

Moreover, the usage of the code has been simplified in several ways. Dependencies from third-party software have been made optional (\rootcern{}) or hidden (\healpix{} is shipped with the code), the compilation process has been improved, and an automated test suite is provided. Also, the input interface has been simplified, facilitating the isolated use of single modules and interfacing the code with other software. Finally, the documentation has also been fully rearranged and updated: it contains a comprehensive description of all features since the first release, step-by-step tutorials and usage examples, available output formats and transformations, as well as background information of numerical implementations and basic concepts. Also, we provide examples of how the code can be interfaced with \python{}. 

With this third release, we hope to further improve the usage of \clumpy{} as a comprehensive tool for indirect DM searches. Finally, let us stress that we welcome suggestions for new functionalities and incorporating new results from the community; any well-motivated request will be considered such that \clumpy{} may keep evolving in-between major releases.

\section*{Acknowledgements}
This work has been supported by the ``Investissements d'avenir, Labex
ENIGMASS", and by the French ANR, Project DMAstro-LHC,
ANR-12-BS05-0006. The work of M.H. was additionally supported by the Research Training Group 1504 of the German Research Foundation (DFG), DESY, and the Max Planck society (MPG). We also thank Deepak Tiwari for identifying a bug in the neutrino mixing matrix in the previous release of the code.

\bibliography{clumpy3}
\bibliographystyle{cpc}
\end{document}

%% file: table_params.tex
\begin{table*}
\caption{\clumpy{} parameters sorted by block (cosmology, dark matter, etc.). New parameters are highlighted in \colorbox{lg}{grey}, deprecated ones in \sout{strikethrough}. Note that this table is not comprehensive; a complete list of parameters is provided with the online documentation at \url{http://lpsc.in2p3.fr/clumpy/}.}
\vspace{-0.2cm}
\begin{center}
{\footnotesize
\begin{tabular}{ll}  \hline\hline
Name & Definition\\ \hline
\multicolumn{2}{l}{{\bf Cosmological parameters} (updated from Planck results)}\\
{\tt gCOSMO\_HUBBLE} & Present-day $(z=0)$ normalized Hubble expansion rate $h=H_0/(100\;{\rm km\;s}^{-1}\;{\rm Mpc}^{-1})$ \\
\sout{{\tt gCOSMO\_RHO0\_C}} & Critical density of the universe, now calculated as $\varrho_{\rm c,0}=3H_0^2/(8\pi G)$ \\[-0.08cm]
{\tt gCOSMO\_OMEGA0\_\{M,\colorbox{lg}{B}\}} & Present-day pressureless matter and baryon-only densities $ \Omega_{\rm m,b} = \varrho_{\rm m,b}/\varrho_{\rm c}$ \\
{\tt \sout{gCOSMO\_OMEGA0\!\_LAMBDA }} & No longer input parameter, now calculated from $\sum_i \Omega_i =1$\\\rowcolor{lg}
{\tt gCOSMO\_SIGMA8} & Fluctuation amplitude, $\sigma_8$, at 8~h$^{-1}$~Mpc \\\rowcolor{lg}
{\tt gCOSMO\_N\_S} & Scalar spectral index $n_{\rm s}$ of primordial fluctuations\\\rowcolor{lg}
{\tt gCOSMO\_T0} & Present-day CMB photon temperature $T_0$ [K] \\\rowcolor{lg}
{\tt gCOSMO\_TAU\_REIO} & Reionization optical depth $\tau_{\rm reio}$ \\\rowcolor{lg}
{\tt gCOSMO\_\{DELTA0, FLAG\_DELTA\_REF\}} & Value $\Delta(z=0)$, and model to describe $\Delta(z)$, see \tab{tab:enum} for possible keywords\\
&\\[-0.2cm]
\multicolumn{2}{l}{{\bf Generic dark matter parameters} (only new, renamed, or deprecated parameters are listed)}\\
{\tt \sout{gDM\_LOGCVIR\_STDDEV}} \colorbox{lg}{\tt gDM\_LOGCDELTA\_STDDEV}  & Width of log-normal $c_\Delta(M_\Delta)$ distribution (value 0 corresponds to a Dirac distribution)\\[-0.08cm]
{\tt \sout{gDM\_FLAG\_CVIR\_DIST}}        & Redundant with {\tt gDM\_LOGCDELTA\_STDDEV} \\[-0.05cm]
{\tt \sout{gDM\_MMIN\_SUBS}} \colorbox{lg}{\tt gDM\_SUBS\_MMIN}              & Minimal mass of DM subhaloes [$M_\odot$] \\[-0.12cm]
{\tt \sout{gDM\_MMAXFRAC\_SUBS}} \colorbox{lg}{\tt gDM\_SUBS\_MMAXFRAC}          & Maximal mass of subhalos in the host halo: $m_{\rm max} =~${\tt gDM\_SUBS\_MMAXFRAC}\;$\times\;M_{\rm host}$\\\rowcolor{lg}
{\tt gDM\_IS\_IDM}    & Use halo mass function cut-off for interacting DM according to \cite{2016JCAP...08..069M}\\\rowcolor{lg}
{\tt gDM\_KMAX}    & Modify scale $k_{\rm max}$ above which $P_{\rm lin}(k)$ is suppressed\\
&\\[-0.2cm]
\multicolumn{2}{l}{{\bf Milky-Way DM halo and subhalo structural parameters (used in {\tt -g} module):} {\tt \sout{gGAL}}$\,\,\rightarrow\,\,${\tt \colorbox{lg}{gMW}} for all parameters}\\
{\tt gMW\_TOT\_\{FLAG\_PROFILE, RSCALE\}}  & Description of the total DM density profile (see \tab{tab:enum}) and its scale radius, $r_{\rm s}$ [kpc]\\[-0.07cm]
{\tt gMW\_TOT\_SHAPE\_PARAMS\sout{\{\![0]\!,\![1]\!,\![2]\!\}}\colorbox{lg}{\!\!\_\{0\!,1\!,2\}}} & Shape parameters of the Milky Way total DM density profile\\[-0.08cm]
{\tt gMW\_\{RHOSOL, RSOL, \sout{RVIR}\colorbox{lg}{\!\!RMAX}\}}                 & Local DM density [GeV~cm$^{-3}$], distance GC--Sun [kpc], DM halo radius [kpc]\\[-0.05cm]
{\tt gMW\_TRIAXIAL\_\{IS, AXES\colorbox{lg}{\!\!\_\{0\!,1\!,2\}\!\!}, ROTANGLES\colorbox{lg}{\!\!\_\{0\!,1\!,2\}\!\!}$\,$\}}                    & Parameters related to triaxiality of the Milky Way DM host halo\\[-0.08cm]
{\tt gMW\_\sout{CLUMPS}\colorbox{lg}{\!\!SUBS\!\!}\_\{FLAG\_PROFILE, \sout{CVIRMIR}\colorbox{lg}{\!\!CDELTAMDELTA\!\!}\}}       & Description of the subhalos'  density profile and $c_\Delta(m_\Delta)$, see \tab{tab:enum} for keywords\\[-0.07cm]
{\tt gMW\_\sout{CLUMPS}\colorbox{lg}{\!\!SUBS\!\!}\_SHAPE\_PARAMS\colorbox{lg}{\!\!\_\{0\!,1\!,2\}}} & Shape parameters of the Milky Way's subhalos'  density profile (always spherical)\\[-0.07cm]
{\tt gMW\_\colorbox{lg}{\!\!SUBS\_\!\!}DPDV\_\{FLAG\_PROFILE, RSCALE\colorbox{lg}{\!\!\_TO\_RS\_HOST}\!\!\}} & Spatial subhalo distribution (see \tab{tab:enum}) in the Milky Way halo  and ratio $a$ (see \secref{sec:dpdvscaleradius})\\[-0.07cm]
{\tt gMW\_\colorbox{lg}{\!\!SUBS\_\!\!}DPDV\_SHAPE\_PARAMS\colorbox{lg}{\!\!\_\{0\!,1\!,2\}}} & Shape parameters of the subhalos' spatial distribution profile\\[-0.08cm]
{\tt gMW\_\colorbox{lg}{\!\!SUBS\_\!\!}DPDM\_SLOPE}                            & Log-slope $\alpha$ of the subhalo mass function $\dd \mathcal{P}/\dd m\propto m^{-\alpha}$\\[-0.01cm]
{\tt gMW\_SUBS\_\{M1, M2, N\_INM1M2\}}            & Number of Milky-Way subhaloes in $[m_1,m_2]$ \\
&\\[-0.2cm]
\multicolumn{2}{l}{{\bf Structural parameters for dedicated halo types (mostly used in {\tt -h} module):} For {\tt TYPE = DSPH, GALAXY, CLUSTER, \colorbox{lg}{EXTRAGAL}\!\!}}\\
\sout{{\tt gTYPE\_TRIAXIAL\_\{IS, AXES\{\![0]\!,\![1]\!,\![2]\!\}, ROTANGLES\{\![0]\!,\![1]\!,\![2]\!\}} }                   & Deprecated: Triaxiality parameters of the host halo are set in {\tt gLIST\_HALOES}\\[-0.08cm]
{\tt gTYPE\_\sout{CLUMPS}\colorbox{lg}{\!\!SUBS\!\!}\_\{FLAG\_PROFILE, \sout{CVIRMIR}\colorbox{lg}{\!\!CDELTAMDELTA\!\!}\}}     & Description of subhalos' density profile in host  and  $c_\Delta(m_\Delta)$, see keywords in \tab{tab:enum}\\[-0.07cm]
{\tt gTYPE\_\sout{CLUMPS}\colorbox{lg}{\!\!SUBS\!\!}\_SHAPE\_PARAMS\colorbox{lg}{\!\!\_\{0\!,1\!,2\}}} & Shape parameters of the host halo {\tt TYPE} subhalos' spherical density profile \\[-0.07cm]
{\tt gTYPE\_\colorbox{lg}{\!\!SUBS\_\!\!}DPDV\_\{FLAG\_PROFILE, RSCALE\colorbox{lg}{\!\!\_TO\_RS\_HOST}\}}\!\!\!\! & Spatial subhalo distribution  (see \tab{tab:enum}) in host {\tt TYPE} and ratio $a$ (see \secref{sec:dpdvscaleradius}) \\[-0.07cm]
{\tt gTYPE\_\colorbox{lg}{\!\!SUBS\_\!\!}DPDV\_SHAPE\_PARAMS\colorbox{lg}{\!\!\_\{0\!,1\!,2\}}} & Shape parameters for the subhalos' spatial distribution in the host \\[-0.07cm]
{\tt gTYPE\_\colorbox{lg}{\!\!SUBS\_\!\!}DPDM\_SLOPE}                          &  Log-slope $\alpha$ of the subhalo mass function $\dd \mathcal{P}/\dd m\propto m^{-\alpha}$\\[-0.03cm]
{\tt gTYPE\_SUBS\_MASSFRACTION}                   &  Mass fraction of the host halo {\tt TYPE} contained in subhalos \\
&\\[-0.2cm]
\multicolumn{2}{l}{{\bf Generic extragalactic physics parameters (used in {\tt -e} module)}}\\\rowcolor{lg}
{\tt gEXTRAGAL\_SUBS\_DPDM\_SLOPE\_LIST}        & List of slopes of subhalo mass spectra $\dd \mathcal{P}/\dd M$ (to be used with the {\tt -e2} submodule)\\\rowcolor{lg}
{\tt gEXTRAGAL\_FLAG\_CDELTAMDELTA\_LIST}       & List of  $c_\Delta(m_\Delta)$ models (to be used with the {\tt -e2} submodule)\\\rowcolor{lg}
{\tt gEXTRAGAL\_FLAG\_MASSFUNCTION}             & Halo mass function multiplicity function $f(\sigma,\,z)$,  \eq{eq:halomassfunction}, see \tab{tab:enum} for keywords \\\rowcolor{lg}
{\tt gEXTRAGAL\_IDM\_\{MHALFMODE, ALPHA, BETA, GAMMA, DELTA\}} & Interacting DM halo mass function cut-off parameters, according to Eq.~(11) of \cite{2016JCAP...08..069M}\\\rowcolor{lg}
{\tt gEXTRAGAL\_HMF\_SMALLSCALE\_DPDM\_\{SLOPE, MLIM\}}   & Force halo mass function log-slope to be steeper than value below $M_{\rm lim}$, ($-1=$ off)\\\rowcolor{lg}
{\tt gEXTRAGAL\_FLAG\_ABSORPTIONPROFILE}        & EBL absorption profile  $\tau(E_{\gamma},z)$, see \tab{tab:enum} for possible keywords\\

&\\[-0.2cm]
\multicolumn{2}{l}{{\bf Particle physics \& spectral parameters:} No changes w.r.t. last release, see Tab.~(2) from \cite{2016CoPhC.200..336B} for the comprehensive list.}\\
{\tt gPP\_\ldots}        & \\
&\\[-0.2cm]
\multicolumn{2}{l}{{\bf Statistical analysis (used in {\tt -s} module)}: previously unnamed command-line parameters. Only a few are listed below.}\\\rowcolor{lg}
{\tt gSTAT\_\{CL, CL\_LIST\}}        & Confidence level (or list of confidence levels) for statistical analysis\\\rowcolor{lg}
{\tt gSTAT\_IS\_LOGL\_OR\_CHI2}        & Choose between a likelihood or $\chi^2$-test based analysis \\\rowcolor{lg}
{\tt gSTAT\_\ldots}        & See online documentation for the comprehensive list\\
&\\[-0.2cm]
\multicolumn{2}{l}{{\bf Draw halos from definition file}}\\
{\tt gLIST\_\{HALOES, HALOES\_JEANS\}}        & Text file defining halo properties for $J$-factor calculations or Jeans analysis\\\rowcolor{lg}
{\tt gLIST\_HALONAME}        & Select a halo by name from above lists for analysis or drawing into 2D skymap \\
&\\[-0.2cm]
\multicolumn{2}{l}{{\bf Global simulation parameters:} {\tt \sout{gSIMU}}$\,\,\rightarrow\,\,${\tt \colorbox{lg}{gSIM}} for all parameters of previous release.  Extragalactic and a few other parameters are listed below.}\\\rowcolor{lg}
{\tt gSIM\_\{NX, IS\_XLOG, JFACTOR, REDSHIFT,\:\ldots\:\}} & Set 1D grid, $J/D$, $z$, or other variables previously only accesible via the command line \\\rowcolor{lg}
{\tt gSIM\_\{PSI\_OBS,THETA\_OBS,THETA\_ORTH\_SIZE,THETA\_SIZE\}\_DEG} & Set FOV position and dimensions for 2D runs (previously only via the command line)\\\rowcolor{lg}
{\tt gSIM\_EXTRAGAL\_\{ZMIN, ZMAX, NZ, IS\_ZLOG\}} & Redshift range and grid for 2D analyses in $(M,z)$ plane\\\rowcolor{lg}
{\tt gSIM\_EXTRAGAL\_\{MMIN, MMAX, NM, IS\_MLOG\}}& Mass range and grid for 2D analyses in $(M,z)$ plane\\\rowcolor{lg}
{\tt gSIM\_EXTRAGAL\_\{DELTAZ, NM\}\_PRECOMP}     & Resolution of redshift (lin) and mass (log) nodes for precalculated integration grid \\\rowcolor{lg}
{\tt gSIM\_EXTRAGAL\_FLAG\_WINDOWFUNC}         & Spherical collapse model window function. See \tab{tab:enum} for possible keywords\\\rowcolor{lg}
{\tt gSIM\_EXTRAGAL\_EBL\_UNCERTAINTY}         & Systematic uncertainty on EBL extinction $\tau(z,E)$\\\rowcolor{lg}
{\tt gSIM\_EXTRAGAL\_FLAG\_GROWTHFACTOR}       & Method to compute perturbation growth factor $g(z)$, \eq{eq:growth}, see \tab{tab:enum} for keywords\\\rowcolor{lg}
{\tt gSIM\_\ldots}       & See online documentation for the comprehensive list\\
%
%
\hline
\end{tabular}
}
\label{tab:params}
\end{center}
\vspace{-0.5cm}
\end{table*}

%% file: table_enums.tex
\begin{table*}
\caption{Enumerators and allowed keywords (with references) in the \clumpy{} code. Curly brackets are used to group keywords having different endings.}
\begin{center}
{
\small
\begin{tabular}{ll}  \hline
Enumerator & Available keywords\\\hline
\multicolumn{2}{c}{\em \colorbox{lg}{New} keywords in this release}\\[0.1cm]\rowcolor{lg}
{\tt gENUM\_ABSORPTIONPROFILE}\!\!\!  & {\tt kFRANCESCHINI\{08, 17\}\,\cite{2008A&A...487..837F,2017A&A...603A..34F}\!, kFINKE10\,\cite{2010ApJ...712..238F}\!, kDOMINGUEZ11\_\{REF, LO, UP\}\,\cite{2011MNRAS.410.2556D}\!,}\\[-0.05cm]\rowcolor{lg}
                                & {\tt kGILMORE12\_\{FIDUCIAL, FIXED\}\,\cite{2012MNRAS.422.3189G}\!, INOUE13\_\{REF, LO, UP\}\,\cite{2013ApJ...768..197I}}\\[-0.05cm]\rowcolor{lg}
{\tt gENUM\_DELTA\_REF}         & {\tt kRHO\_CRIT, kRHO\_MEAN, kBRYANNORMAN98\,\cite{1998ApJ...495...80B}}\\[-0.05cm]\rowcolor{lg}
{\tt gENUM\_GROWTHFACTOR}       & {\tt kHEATH77\,\cite{2010ARA&A..48..673F}\!, kCARROLL92\,\cite{1992ARA&A..30..499C}\!, kPKZ\_FROMFILE$^\dagger$}\\[-0.05cm]\rowcolor{lg}
{\tt gENUM\_MASSFUNCTION}       & {\tt kPRESSSCHECHTER74\,\cite{1974ApJ...187..425P}\!, kSHETHTORMEN99\,\cite{1999MNRAS.308..119S}\!, kJENKINS01\,\cite{2001MNRAS.321..372J}\!,}\\[-0.05cm]\rowcolor{lg}
                                & {\tt kTINKER\{08, 08\_N, 10\}\,\cite{2008ApJ...688..709T,2008ApJ...688..709T,2010ApJ...724..878T}\!, kBOCQUET16\_\{HYDRO, DMONLY\}\,\cite{2016MNRAS.456.2361B}\!,}\\[-0.05cm]\rowcolor{lg}
                                & {\tt kRODRIGUEZPUEBLA16\_PLANCK\,\cite{2016MNRAS.462..893R}}\\ [-0.05cm]\rowcolor{lg}
{\tt gENUM\_WINDOWFUNC}         & {\tt kTOP\_HAT, kGAUSS, kSHARP\_K}\\
                                & \\[-0.1cm]
\multicolumn{2}{c}{{\em Keywords already present in v1 and v2 (new keywords} \colorbox{lg}{highlighed})}\\[0.1cm]
{\tt gENUM\_ANISOTROPYPROFILE}\!\!\!  & {\tt kCONSTANT, kBAES\,\cite{2007A&A...471..419B}\!, kOSIPKOV\,\cite{1979PAZh....5...77O,1985AJ.....90.1027M}}\\[-0.05cm]
{\tt gENUM\_LIGHTPROFILE}       & {\tt kEXP2D\,\cite{2009MNRAS.393L..50E}\!,\,kEXP3D\,\cite{2009MNRAS.393L..50E}\!,\,kKING2D\,\cite{1962AJ.....67..471K}\!,\,kPLUMMER2D\,\cite{1911MNRAS..71..460P}\!,\,kSERSIC2D\,\cite{1968adga.book.....S}\!,\,kZHAO3D\,\cite{1990ApJ...356..359H,1996MNRAS.278..488Z}}\!\!\!\!\\[-0.05cm]
%
{\tt gENUM\_\colorbox{lg}{CDELTAMDELTA}}$^\S$& {\tt kB01\_\{VIR, \colorbox{lg}{\!VIR\_RAD\!}\}\,\cite{2001MNRAS.321..559B,2008ApJ...686..262K}\!, kENS01\_VIR\,\cite{2001ApJ...554..114E}\!, kNETO07\_200\,\cite{2007MNRAS.381.1450N}\!,}\\[-0.05cm]
                                & {\tt kDUFFY08F\_\{VIR,200,MEAN\}\,\cite{2008MNRAS.390L..64D}\!,\,kETTORI10\_200\,\cite{2010A&A...524A..68E}\!,\,kPRADA12\_200\,\cite{2012MNRAS.423.3018P}\!,\,kGIOCOLI12\_VIR\,\cite{2012MNRAS.422..185G}}\\[-0.05cm]
                                & {\tt kPIERI11\_\{VIALACTEA,AQUARIUS\}\,\cite{2011PhRvD..83b3518P}\!,\,\colorbox{lg}{kROCHA13\_SIDM\_VIR\,\cite{2013MNRAS.430...81R}}\!,\,kSANCHEZ14\_200\,\cite{2014MNRAS.442.2271S}\!},\\[-0.05cm]
                                & \!\!{\tt \colorbox{lg}{kCORREA15\_PLANCK\_200\,\cite{2015MNRAS.452.1217C}\!,\,kLUDLOW16\_200\,\cite{2016MNRAS.460.1214L}\!,\,kMOLINE17\_200\,\cite{2017MNRAS.466.4974M}}}\\[-0.05cm]
%
{\tt gENUM\_PROFILE}            & {\tt kHOST,\,kZHAO\,\cite{1990ApJ...356..359H,1996MNRAS.278..488Z}\!,\,kEINASTO\,\cite{2004MNRAS.349.1039N}\!,\,kEINASTO\_N\,\cite{2006AJ....132.2685M}\!,\,kBURKERT\,\cite{1995ApJ...447L..25B}\!,\,\colorbox{lg}{kISHIYAMA14}\,\cite{2014ApJ...788...27I}\!,}\\[-0.05cm]
                                & {\tt \sout{kGAO\_SUB}\colorbox{lg}{kDPDV\_GAO04}$^\ddagger$\,\cite{2004MNRAS.355..819G,2008ApJ...679.1260M}\!,\colorbox{lg}{kDPDV\_SPRINGEL08\_FIT}$^\ddagger$\,\cite{2008MNRAS.391.1685S}\!,}\\[-0.05cm]
                                & {\tt\sout{kEINASTOANTIBIASED\_SUB}\colorbox{lg}{kDPDV\_SPRINGEL08\_ANTIBIASED}$^\ddagger$\,\cite{2007ApJ...671.1135K,2015MNRAS.447..939L}}\\[-0.05cm]
{\tt gENUM\_TYPEHALOES}         & {\tt kDSPH, kGALAXY, kCLUSTER, \colorbox{lg}{kEXTRAGAL}}\\[-0.05cm]
{\tt gENUM\_FINALSTATE}         & {\tt kGAMMA, kNEUTRINO}\\[-0.05cm]
{\tt gENUM\_NUFLAVOUR}          & {\tt kNUE, kNUMU, kNUTAU}\\[-0.05cm]
{\tt gENUM\_PP\_SPECTRUMMODEL}\!\!\!\!\!\!  & {\tt kBERGSTROM98$^\star$\,\cite{1998APh.....9..137B}\!, kTASITSIOMI02$^\star$\,\cite{2002PhRvD..66h3006T}\!, kBRINGMANN08$^\star$\,\cite{2008JHEP...01..049B}, kCIRELLI11\_\{EW, NOEW\}\,\cite{2011JCAP...03..051C}}\\
\hline\hline
\end{tabular}
}
\label{tab:enum}
{\footnotesize
\\$^\dagger$ Look for files in {\tt data/pk\_precomp}, and if absent for user cosmology, generate $P(k,z)$ files with \texttt{CLASS} code \cite{2011arXiv1104.2932L}.
\\$^\S$ To be consistant with {\tt gENUM\_DELTA\_REF} notation, we renamed {\tt gENUM\_CVIRMVIR} to {\tt gENUM\_CDELTAMDELTA}.
\\$^\ddagger$ Applicable for the spatial distribution of subhaloes only.
\\$^\star$ For $\gamma$-rays only (not enabled for $\nu$).
\\
}
\end{center}
\vspace{-0.3cm}
\end{table*}